\documentclass[12pt]{iopart}
\pdfoutput=1

\usepackage{amssymb}
\usepackage{graphicx}
\usepackage{dcolumn}
\usepackage{bm}
\usepackage{psfrag}
\usepackage{bbold}
\usepackage{amsthm}
\usepackage{iopams}
\usepackage{color}

\newcommand{\be}{\begin{equation}}
\newcommand{\ee}{\end{equation}}
\newcommand{\sysb}{\left\{\begin{array}}
\newcommand{\syse}{\end{array}\right.}
\newcommand{\baa}{\begin{array}}
\newcommand{\eaa}{\end{array}}

\newcommand{\matb}{\left(\begin{array}}
\newcommand{\mate}{\end{array}\right)}

\newcommand{\lan}{\left\langle}
\newcommand{\ran}{\right\rangle}
\newcommand{\ket}[1]{\left| #1 \ran}
\newcommand{\bra}[1]{\lan #1 \right|}

\begin{document}

\title[Crystalline structures]{Crystalline structures in a one-dimensional two-component lattice gas with $1/r^{\alpha}$ interactions}

\author{Emanuele Levi, Ji\v{r}\'i Min\'a\v{r} and Igor Lesanovsky}
\address{School of Physics and Astronomy, The University of Nottingham, Nottingham, NG7 2RD, United Kingdom}

\begin{abstract}
We investigate the ground state of a one-dimensional lattice system that hosts two different kinds of excitations (species) which interact with a power-law potential. Interactions are only present between excitations of the same kind and the interaction strength can be species-dependent. For the case in which only one excitation is permitted per site we derive a prescription for determining the ground state configuration as a function of the filling fractions of the two species. We show that depending on the filling fractions compatible or incompatible phases emerge. Furthermore, we discuss in detail the case in which one species is strongly and the other one weakly interacting. In this case the configuration of the strongly interacting (strong) species can be considered frozen and forms an effective inhomogeneous lattice for the other (weak) species. In this limit we work out in detail the microscopic ground state configuration and show that by varying the density of the weak species a series of compatible--incompatible transitions occurs. Finally we determine the stability regions of the weak species in the compatible phase and compare it with numerical simulations.
\end{abstract}
\maketitle

\section{Introduction}

Understanding the ground state properties of one-dimensional lattice gases or spin-models with long--range interaction is a notoriously difficult problem in many--body physics. Only few analytical solutions for classical \cite{Hubbard,Pokrovsky} and quantum \cite{Haldane,Shastry,Hikami,Levi1} models have so far been found. Moreover, quantum systems with long-range interactions are typically difficult to treat numerically even with modern numerical techniques based on the density-matrix renormalization group \cite{White1,Schwollok,Perez,Verstraete}.

Despite these difficulties physical systems with power-law interactions have received significant attention in the last decades.
In early years the the Kondo problem for the one-dimensional Ising model with inverse square interactions \cite{algebraicKondo}, and the phase diagram thereof \cite{Cardy} were analyzed with renormalization group techniques.
In more recent years the critical properties of spin models with general algebraic interactions and dimensionality were studied \cite{Luijten1996,Luijten2001,Luijten2002}.

This work connects to the studies in Refs. \cite{Hubbard,Pokrovsky} where a classical one-dimensional Ising model with an external magnetic field and convex (long-ranged) interactions was investigated, for which the ground state was constructed analytically. In Ref. \cite{Bak} it was moreover shown that here the 'spin-up' excitations, upon varying their density, form a complete devil's staircase. The role of quantum fluctuations in these spin models was explored for the case of power-law potentials of the form $1/r^\alpha$ in the context of dipolar Bose gases \cite{Burnell} ($\alpha=3$) and strongly interacting Rydberg gases \cite{Weimer} ($\alpha=6$). Recently, the onset of a devil's staircase has indeed been identified in a one-dimensional lattice gas of atoms that were laser-excited to electronically high-lying Rydberg states \cite{Bloch}.

In this paper we generalize the aforementioned classical studies to a
two-component lattice system. Specifically, we consider a situation in
which there are two different kinds of excitations to which we will
refer as $s$ (strong) and $w$ (weak). Power-law interactions among
$s$($w$)-excitations are assumed to be strong (weak) while there is
negligible interaction between excitations belonging to different
components. One motivation for conducting this study is to show that the
physics of this system is much richer than the single-component case and
in fact features a series of transitions that are absent in the
single-component situation. A second motivation is derived from the fact that multi-component lattice gases with power-law interactions move more and more into the focus of theoretical and experimental studies on cold atomic gases in which atoms are excited to Rydberg states
\cite{Ditzhuijzen08,gunter_observing_2013,Bettelli13,Maxwell14,Li14,Gorniaczyk14,Barredo14,Teixeira}.
In certain regimes aspects of the physics of this atomic system ---
specifically, strongly state-dependent interactions and highly
suppressed interactions between atoms in different states \cite{Teixeira,Olmos}  --- are indeed  well described by the model
discussed here.

The paper is organized as follows. In Section \ref{sec:model} we define the model. We then summarize the results obtained in \cite{Hubbard,Pokrovsky,Bak} in Section \ref{sec:single} for the single-species case. We do so because we will make an extensive use of the technology and formalism developed in those works. In Section \ref{sec:comm} we introduce the concept of \textit{compatible} and \textit{incompatible} ground states and in Section \ref{sec:disposition} we introduce an algorithm for finding the microscopic arrangement of excitations in the ground state.
Finally in Section \ref{sec:stability} we identify the locations of the compatible-incompatible transitions in the special case of filling fractions of the form $1/q$, $q \in \mathbb{N}^+$.

\section{The model}
\label{sec:model}

We focus on a one--dimensional lattice system where each site is occupied by a three--level system whose internal states belong to the set $\left\{\ket{0},\ket{s},\ket{w}\right\}$.
The three levels correspond to the (empty) state $0$ and the cases in which the site is occupied by an $s$- or $w$-excitation. Excitations of the same kind interact with an inverse power-law potential with exponent $\alpha$ and strength $V_{\mu}$ (with $\mu=s,w$ and $V_{\mu}>0$), while there is no interaction between species of different type. The Hamiltonian then reads
\begin{equation}
  \label{eq:ham2spec}
 H=\sum_{k=1}^N \sum_{\mu=s,w}\left(-h_{\mu} n_k^{(\mu)}+V_{\mu} \sum_{l > k}\frac{n_k^{(\mu)}n_l^{(\mu)}}{(l-k)^{\alpha}} \right),
\end{equation}
where $n^{(\mu)}=\ket{\mu}\bra{\mu}$ and $h_{\mu}$ is a positive parameter. Linking this model to experiments with gases of Rydberg atoms, one would think of $\ket{0}$ as being the atomic ground state. The state $\ket{\mu}$ then corresponds to an electronically high-lying Rydberg $\nu S$-state with principal quantum number $\nu_\mu$ \cite{Rydberg} which is excited with a laser of detuning $h_\mu$. In order to have one strongly and one weakly interacting species one needs to require $\nu_s \gg \nu_w$\footnote{In fact the strength of $V_\mu$ among excitations of the same species (atoms in the same Rydberg state) scales with the eleventh power of the principal quantum number $\nu_\mu$. Moreover, as shown by a perturbative calculation \cite{Olmos} and recently also experimentally \cite{Teixeira} there are states for which the interactions among excitation of different species is strongly suppressed.}.
In the following, unless specified otherwise, we consider the asymptotic scenario $V_s/V_w \rightarrow \infty$.

Our goal is to find the configuration of the two species that minimizes the energy of Hamiltonian (\ref{eq:ham2spec}) given fixed values of the individual total excitation numbers $n_{\mu}=\sum_{k=1}^N n_k^{(\mu)}$. The condition $V_s/V_w \rightarrow \infty$ renders the search for the ground state into a two-step process: First, the ground state configuration of the $s$-excitations is to be found which subsequently remains ``frozen''. This leads to an effective --- in general inhomogeneous --- lattice on which excitations of the $w$-species are to be arranged in the second step. Depending on the imposed total excitation numbers $n_\mu$ one can distinguish dispositions of the $w$-excitations that are \textit{compatible} with the one of the $s$-excitations, and others that are \textit{incompatible}.
Transitions between these two cases can then be studied as a function of the total excitation numbers $n_\mu$ (see e.g. \cite{Compatible} for a related discussion of compatible-incompatible transitions).

\section{The ground state of the single-species problem}
\label{sec:single}
 \begin{figure}
 \begin{center}
\includegraphics[width=.5\textwidth]{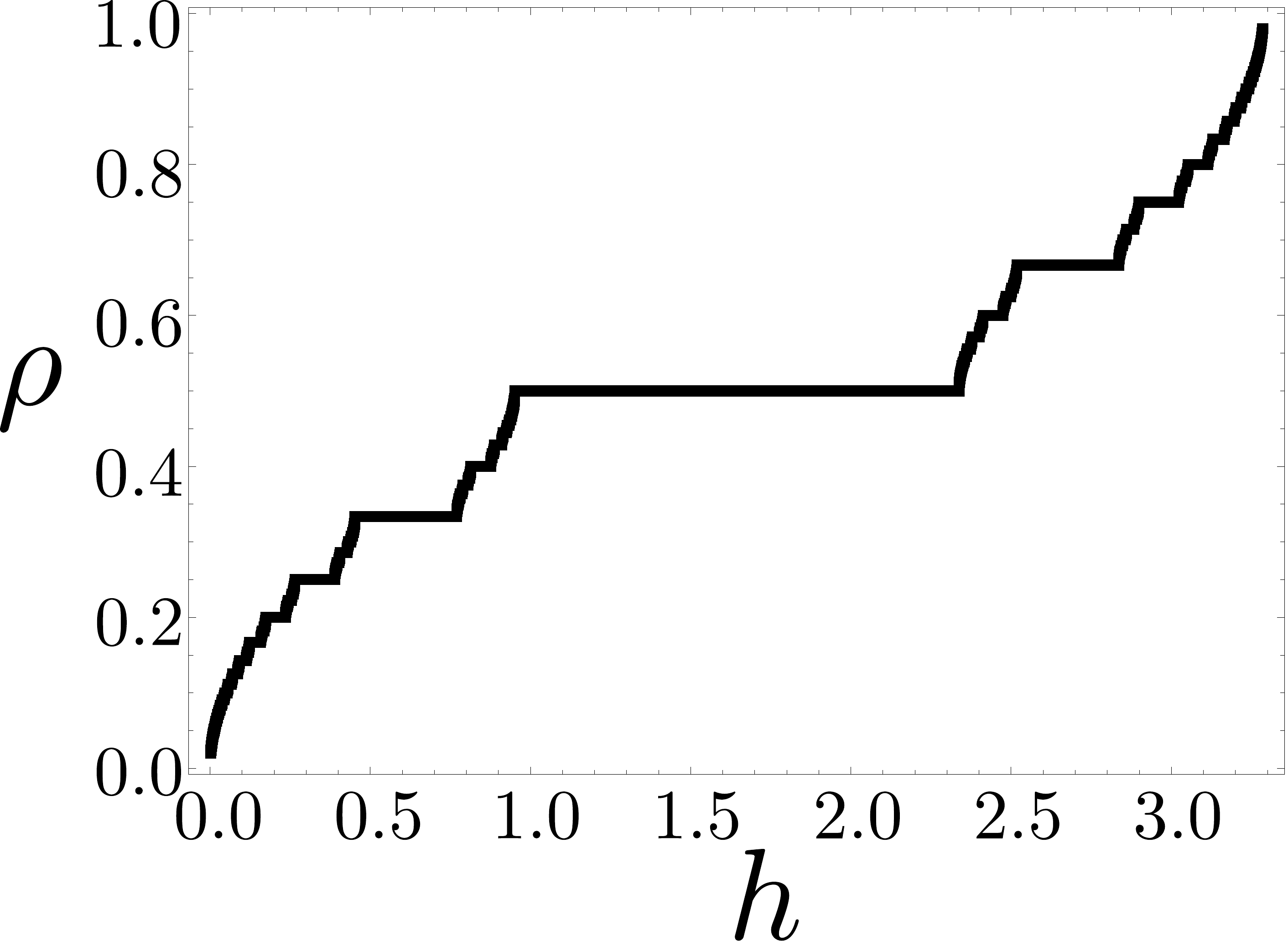}
\end{center}
  \caption{The single species devil's staircase for a potential $V(r)=1/r^2$.}
  \label{fig:devil}
 \end{figure}
The single-species case can be described in terms of the Hamiltonian (\ref{eq:ham2spec}) once we set e.g. $V_w=h_w=0$. Our aim is to find the configuration that minimizes the interaction energy once a density of excitations $\rho=m/N$ is given\footnote{In this section we drop the label $s$ as we consider a single species.}. For a start let us consider three excitations arranged on a line with distances $r$ (between the first and the second one) and $r'$ (between the second and the third one) with $r<r'$. The solution of the energy minimization problem is obtained by realizing that for a general convex potential $V(r)$ the variation of the potential energy under a change of the position of the second excitation by $\delta r$ obeys
\begin{equation}
 \label{eq:convex}
 V(r+\delta r)+V(r'-\delta r)\leq V(r)+V(r').
\end{equation}
This expression holds for $r,\delta r<r'$. Hence moving the excitation in the middle towards the third excitation leads to a lowering of the potential energy until $r=r'$. The energetically most favourable configuration is therefore assumed when both distances are equal, i.e. the excitations are arranged in the most uniform distribution. This consideration can be extended to larger systems with excitation density $\rho$ where the ground state is a regular arrangement of excitations with inter-excitation distance $1/\rho$.

On a chain, however, where the coordinates can only be multiples of the lattice spacing the situation is much less trivial. Here the construction of the ground state configuration proceeds as follows (Note, that the lattice spacing is set to one, throughout.): We define $x_k$ as the position of the $k$-th excitation and denote the distance between the $k$-th excitation and its nearest excitation by $r^{(k)}_1=x_{k+1}-x_k$, the distance between the $k$-th and the next-to-nearest excitation by $r^{(k)}_2=x_{k+2}-x_k=r^{(k)}_1+r^{(k+1)}_1$, and generally the distance to the $l$-th-nearest excitation by $r^{(k)}_l=\sum_{j=0}^{l-1}r^{(k+j)}_1=x_{k+l}-x_k$. All spacings of neighboring excitations, $r^{(k)}_1$, will belong to the set $\mathcal{S}_1=\{ \lfloor1/\rho\rfloor, \lceil1/\rho\rceil \}$ which is referred to as the \textit{minimal set}.

Using the above abbreviations the task is now to minimize
\begin{equation}
 \label{eq:minpot}
 E=-mh+\sum_{l=1}^{m}\left[\sum_{k=1}^m V\left(r_l^{(k)}\right)\right].
\end{equation}
In Ref. \cite{Hubbard} it was proposed to look for a solution which minimizes each inner sum separately, knowing that if such solution exists it will also minimize $E$. There it was demonstrated that this solution exists and that it is achieved when $r_l^{(k)}$ form minimal sets, for each $l$. These sets can be defined as $\mathcal{S}_l=\{ r_l,r_l+1 \}$, where $r_l=\lfloor l/\rho\rfloor$, and $r_l+1=\lceil l/\rho\rceil$.

Another challenge is finding the microscopic arrangements of the excitations on an infinite chain with rational filling fraction $\rho=p/q$. Clearly for all the filling fractions of the form $p=1$, $q\in \mathbb{N}^+$ one just has to distribute the excitations uniformly, one every $q$ sites. For general rational numbers it is more complicated to figure out how the excitations are distributed, but we can still think that the configuration will be periodic in $q$, each period having $p$ excitations arranged in the most uniform way.
The algorithm for finding such a distribution is given in Refs. \cite{Hubbard,Pokrovsky}. To illustrate it we start with an example and then report the general solution.

\subsection{Example}
Let us consider the case $\rho=11/47$. From the previous explanation we know that the distances between neighboring excitations are $r_1^{(k)}\in \{4,5\}$. This automatically minimizes $\sum_k V(r_1^{(k)})$. If we just had to minimize this contribution we could distribute the $11$ excitations over the $47$ sites in any pattern, which would ensure that excitations are four or five sites apart. However, as we have to minimize each contribution $l$ in (\ref{eq:minpot}) separately, we need a more sophisticated way to understand the configuration. We start by writing
\begin{equation}
 \label{eq:1/rho}
 \frac{1}{\rho}=4+\frac{3}{11},
\end{equation}
hence finding $r_1^{(k)}\in \mathcal{S}_1=\{4,5\}$. With 47 sites we therefore find 8 intervals $r^{(k)}_1$ of 4 sites, and three of 5 sites. The only possible values for the distances between next-neighboring excitations $r^{(k)}_2$, are then given by 8, 9, 10. Note, that for some part of the following discussion it turns out that it is actually convenient to express these distances as strings of consecutive spacings, i.e. 44, 45, 55. Let us now find the permitted next-nearest-neighbor distances which form the set $\mathcal{S}_2$. This set must contain only two consecutive numbers, so either $8,9$ or $9,10$. In order to understand which ones to pick and how to distribute intervals of length 4 and 5 we define a new filling fraction using the non-integer remainder of (\ref{eq:1/rho}):
\begin{equation}
 \label{eq:1/rho2}
 \frac{11}{3}=4-\frac{1}{3}.
\end{equation}
This filling fraction represents the density of $5$ sites intervals which can be regarded as defects within the $4$ sites intervals. We thus have two strings of length $4$, and one of length $3$ between these defects (4+4+3=11), where the strings of length 4 and 3 correspond to $4445$ and $445$, respectively.

In order to understand how these strings are distributed we iterate the logic and obtain from the remainder of (\ref{eq:1/rho2}) the filling fraction $(1/3)^{-1}=3$. This means that if we identify the string $445$ as the defect, and $4445$ as the defect-free one we have eventually $44454445445$. This is the ground state configuration as the iterative procedure stops when the effective $1/\rho \in \mathbb{N}^+$ (here $(1/3)^{-1} \in \mathbb{N}^+$). This method creates the configuration which minimizes each contribution in (\ref{eq:minpot}) separately. In fact one can check that for $\forall l\leq m$ , $r_l\in \mathcal{S}_l$.

\subsection{General algorithm}
The generalization of this method was given in Ref. \cite{Hubbard}. It starts by considering the set of numbers
\begin{eqnarray}
 \label{eq:hubbmethod}
  \frac{1}{\rho}&=n+r_0\nonumber\\
\left|\frac{1}{r_0}\right|&=n_1+r_1 \nonumber\\
\left|\frac{1}{r_1}\right|&=n_2+r_2\nonumber\\
&\vdots\nonumber\\
\left|\frac{1}{r_{k-2}}\right|&=n_{k-1}+r_{k-1}\nonumber\\
\left|\frac{1}{r_{k-1}}\right|&=n_k,
\end{eqnarray}
where $n_j \in \mathbb{Z}$ and $k$ is a finite number since $\rho\in\mathbb{Q}$, and $-1/2\leq r_j \leq 1/2$ for any $j$. Then one can define iteratively the strings of non-defective sites $X_1,X_2,...,X_k$ and defects $Y_1,Y_2,...,Y_k$ in the following way
 \begin{eqnarray}
  \label{eq:hubbmethod2}
   X_1=n, \hspace{1cm}&Y_1=n+\mathrm{sgn}\left(r_0\right)\nonumber\\
 &\vdots\nonumber\\
 X_{i+1}=\left[X_i\right]^{n_i-1}Y_i, \hspace{1cm}&Y_{i+1}=\left[X_i\right]^{n_i+\mathrm{sgn}\left(r_i\right)-1}Y_i.
 \end{eqnarray}
Here the notation $\left[X\right]^{n}$ means that the spacing $X$ is repeated $n$ times. The ground state configuration is then given by $X_k$.

\subsection{Stability of the ground state}
Finally, we summarize the solution to the problem of understanding in which region of the parameter space --- and in particular for which values of the parameter $h$ --- a given configuration of density $\rho=m/N$ is stable. The solution was given by Bak in Ref. \cite{Bak} assuming a chain of length $N$ with periodic boundary conditions. The essence is to analyze the energy cost of inserting or extracting one excitation from a given configuration. If both operations increase the energy (\ref{eq:minpot}) the configuration is stable.

Adding one excitation will modify the number of $l$-th nearest neighbor distances $r_l$ and $r_l+1$. Starting from a given configuration with $m-a$ distances $r_l$ and $a$ distances $r_l+1$ the following system of equations must hold
 \begin{equation}
  \label{eq:system2}
  \left\{
   \begin{array}{l}
     \sum_{k=1}^m r_l^{(k)}=(m-a)r_l+a(r_l+1)=Nl\\
     \sum_{k=1}^{m+1} r_l^{(k)}=(m+1-a')r_l+a'(r_l+1)=Nl\\
   \end{array}
   \right. .
 \end{equation}
Solving (\ref{eq:system2}) gives the number $a'$ of distances $r_l+1$ in the configuration with one added excitation, $a'=a-r_l$. It means that $r_l$ distances $r_l+1$ will be replaced by $r_l+1$ distances $r_l$. Note that in the special case $m r_l = Nl$, one instead has to replace $r_l-1$ distances $r_l$ by $r_l$ distances $r_l-1$. With the same logic one understands that extracting one excitation from the system will replace $r_l+1$ distances $r_l$ with $r_l$ distances $r_l+1$. Considering the energy (\ref{eq:minpot}) we have
 \begin{eqnarray}
  \label{eq:sdifferences}
     \hspace*{-2cm} \Delta E_+=-h_++(r_1+1)V(r_1)-r_1V(r_1+1)+(r_2+1)V(r_2)-r_2V(r_2+1)+...+\nonumber\\
     \hspace*{-2cm} \phantom{\Delta E_+=}  +qV(q-1)-(q-1)V(q)+...+2qV(2q-1)-(2q-1)V(2q)+...\nonumber\\
     \nonumber\\
     \hspace*{-2cm} \Delta E_-=h_--(r_1+1)V(r_1)+r_1V(r_1+1)-(r_2+1)V(r_2)+r_2V(r_2+1)+...+\nonumber\\
     \hspace*{-2cm} \phantom{\Delta E_-} -(q+1)V(q)+qV(q+1)+...-(2q+1)V(2q)+2qV(2q+1)+...\nonumber\\
 \end{eqnarray}
 where we are denoting with $\Delta E_\pm$ the change in energy when adding or extracting an excitation.
 Setting the two quantities $\Delta E_\pm$ equal to zero gives a region of stability of the width
 \begin{equation}
  \label{eq:region}
  h_+-h_-=\sum_{n}nq\left[V(nq-1)+V(nq+1)-2V(nq)\right].
 \end{equation}
 In \cite{Bak} it was proven that these intervals of stability for rational values of $\rho$ fill up the whole parameter space $h\in\mathbb{R}^{+}$, giving rise to a so called \textit{devil's staircase} as shown in Fig.\ref{fig:devil}.

\section{The two-species problem --- Compatible and incompatible ground state dispositions}
\label{sec:comm}
 \begin{figure}
 \begin{center}
\includegraphics[width=\textwidth]{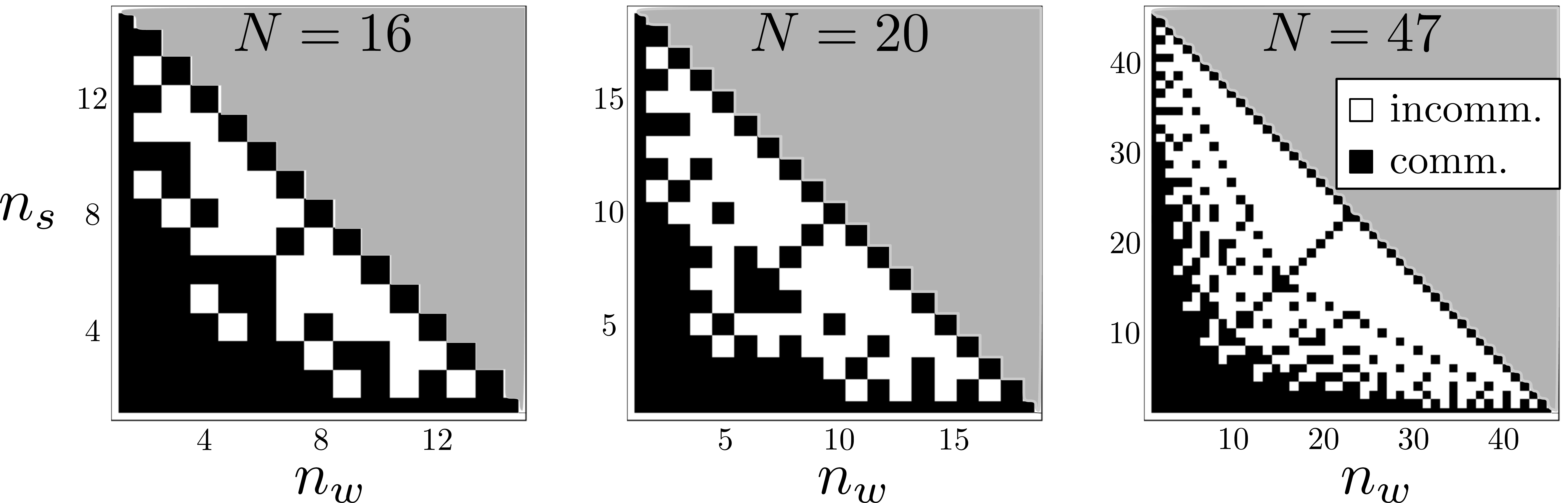}
\end{center}
\caption{Compatible (black squares) and incompatible (white squares) filling fractions for a periodic lattice of length $N=16,20,47$. The results are given for different excitation numbers $n_s$ and $n_w$ corresponding to filling fractions $\rho_s=n_s/N$ and $\rho_w=n_w/N$. The shaded area corresponds to cases $\rho_w+\rho_s>1$ which cannot occur since each lattice site only contains at most a single excitation.} \label{fig:comm}
 \end{figure}

We will now turn to the case in which there are two different species --- $s$ and $w$ --- in the lattice. If we allowed for multiple occupations on each site we would clearly have that the energy (\ref{eq:ham2spec}) is minimized by disposing $s$ and $w$ excitations independently, using the algorithms (\ref{eq:hubbmethod}) and (\ref{eq:hubbmethod2}). However, as we allow just for one excitation per site we have the additional restriction $\rho_s+\rho_w\leq1$. Nevertheless, we will see, that in some cases, i.e. for some filling fractions, the two populations do not influence each other. The more interesting case, however, is encountered when both species cannot simultaneously assume their ground state configuration. We will refer to the former case as \textit{compatible} and to the latter as \textit{incompatible}.

The goal of this section is to define a method for distinguishing the compatible cases from the incompatible ones. Note that this is a purely combinatorial problem and that energy considerations are irrelevant for the notion of compatibility. To distinguish compatible from incompatible cases we consider the filling fractions $\rho_s=n_s/N$ and $\rho_w=n_w/N$, such that the distributions of the $s$- and $w$-species are periodic with the same period $N$. This causes no loss of generality as the filling fractions with two different periods, e.g. $\rho_s=n_s/N_s$ and $\rho_w=n_w/N_w$, can be always expressed in terms of a single period, $\rho_s=n_s N_w/(N_sN_w)$ and $\rho_w=n_w N_s/(N_sN_w)$. The two configurations corresponding to $\rho_s$ and $\rho_w$ are compatible if within a period each excitation of the kind $w$ is located in a hole with respect to the arrangement of the $s$-excitations ($s$-holes). The density of $s$-holes, denoted by $\tilde{\rho}_s$, is linked to the density of $s$-excitations by $\tilde{\rho}_s=1-\rho_s$, and the disposition of such holes can be evaluated by using the algorithm (\ref{eq:hubbmethod2}). As in the previous section we start with an example and then provide the general rule for distinguishing a compatible case from an incompatible one.

\subsection{Example}
Consider the filling fractions $\rho_w=3/16$ and $\rho_s=9/16$. The latter yields $\tilde{\rho}_s=7/16$. The disposition of $w$-excitations is 556, while the disposition of the $s$-holes is 3223222. If one is able to match the positions of the 3 $w$-excitations with the positions of three of the 7 $s$-holes without distorting the excitation distance pattern 556, then the $w$ and $s$ configurations are compatible. One can therefore think of the hole configuration as an effective inhomogeneous or distorted lattice with spacings $\tilde{a}=\{2,3\}$, disposed with repeated periods $...PPP...$, $P=3223222$. Using (\ref{eq:hubbmethod2}) with the effective filling fraction $\rho_w/\tilde{\rho}_s=3/7$, one finds that the configuration of the 3 $w$- excitations on the distorted lattice is $223$. Here the first 2 in the string $223$ means the first two elements in $P$ and so on, i.e. disposing the $w$-excitations on the distorted lattice gives the disposition $223 \rightarrow (32)(23)(222) = 556$. This is identical to the original disposition, so that the dispositions of the $w$- and $s$-species are compatible. With the same logic one finds that the filling fractions $\rho_w=3/16$, and $\rho_s=12/16$ are not compatible. In this case $\tilde{\rho}_s=4/16$, with a configuration of holes 4444, and there is no way we can combine multiples of 4 in a way to match the distances 5 and 6 between $w$-excitations.

\subsection{General algorithm}
The general algorithm to check the compatibility of two distributions of excitations is based on the following sets of numbers
\begin{eqnarray}
\label{eq:commdens}
 \mathcal{P}_l=\left\{r_l,r_l+1\right\}=\left\{\left\lfloor\frac{l}{\rho_w}\right\rfloor ,\left\lceil\frac{l}{\rho_w}\right\rceil \right\},\nonumber\\
 \mathcal{H}_l=\left\{s_l,s_l+1\right\}=\left\{\left\lfloor\frac{l}{\tilde{\rho}_s}\right\rfloor ,\left\lceil\frac{l}{\tilde{\rho}_s}\right\rceil \right\},\nonumber\\
 \mathcal{Q}_l=\left\{q_l,q_l+1\right\}=\left\{\left\lfloor\frac{l\tilde{\rho}_s}{\rho_w}\right\rfloor, \left\lceil\frac{l\tilde{\rho}_s}{\rho_w}\right\rceil \right\}.
\end{eqnarray}
These are the distances $\mathcal{P}_l$ between $l$-th-nearest $w$-excitations and the distances $\mathcal{H}_l$ between the $l$-th-nearest $s$-holes. The third set contains the distances between $l$-th-nearest $w$-excitations counted in units of lattice spacings of the distorted lattice. We focus here on the case $\rho_w/\tilde{\rho}_s<1$ (i.e. $\rho_s/\rho_w > 1$). The opposite case is described by simply inverting the roles of excitations and holes.

The distance between $l$-th nearest $w$ excitations is either $q_l$ or $q_{l+1}$ counted in distorted lattice spacings, where a distorted lattice spacing reads either $s_1$ or $s_1+1$ original lattice spacings.
If it is possible to group $q_l$ and $q_l+1$ distorted lattice spacings $s_1$ and $s_1+1$ in a way that they sum up to give the distances $r_l$ and $r_l+1$ then the two densities are compatible.
For example, the compatible case in Fig. \ref{fig:config} has $\rho_s = 1/2$ and $\rho_w=1/4$.
Notice, that in this simple example the sets Eq.(\ref{eq:commdens}) contain only one number and read $\mathcal{P}_l = \{4l\}, \mathcal{Q}_l = \{2l\}$ and for $l=1$ $\mathcal{H}_1 = \{2\}$. Clearly, the $q_l=2l$ distances $s_1=2$ yield the distance $4l=r_l$, i.e. the two densities are compatible.
On the other hand, considering e.g. the first incompatible filling fractions in Fig. \ref{fig:config} $\rho_s=5/16$, $\rho_w=1/4$ one has for $l=1$ $\mathcal{P}_1 = \{4\}, \mathcal{H}_1 = \{1,2\}, \mathcal{Q}_1 = \{2,3\}$, and in particular the distorted lattice is formed by repeated periods  $P=12121212112$.
One can group 3 consecutive distances in $P$ to get $r_1=4$ (namely 112), but it is not possible to obtain 4 from grouping only 2 consecutive distances, so that already at level $l=1$ one can prove the incompatibility.
Formally, this means that two configurations are compatible if
\begin{equation}
 \label{eq:condition}
 \forall l\leq n_w,\hspace{1cm}\mathcal{H}_{q_l}\cap\mathcal{P}_l\neq \emptyset \wedge \mathcal{H}_{q_l+1}\cap\mathcal{P}_l\neq \emptyset.
\end{equation}
In Fig.\ref{fig:comm} we show a diagram illustrating the filling fractions that lead to compatible or incompatible dispositions of excitations for different system sizes.

As a byproduct of Eq. (\ref{eq:condition}) one learns that a sufficient condition for two filling fractions to be compatible is that they can be represented as $\rho_w=n_w/N$, and $\rho_s=1-k\rho_w$, with $1\leq k \leq \lfloor \rho_w^{-1} \rfloor$. In fact when $k=1$ the two configurations are dual (corresponding to the upper left to lower right diagonal in Fig. \ref{fig:comm}), in the sense that $\rho_w = \tilde{\rho}_s$. Increasing $k$ increases the number of holes while the $w$- and $s$-configurations remain compatible.

\section{Disposition of the excitations in the incompatible case}
\label{sec:disposition}
\begin{figure}
 \begin{center}
\includegraphics[width=.5\textwidth]{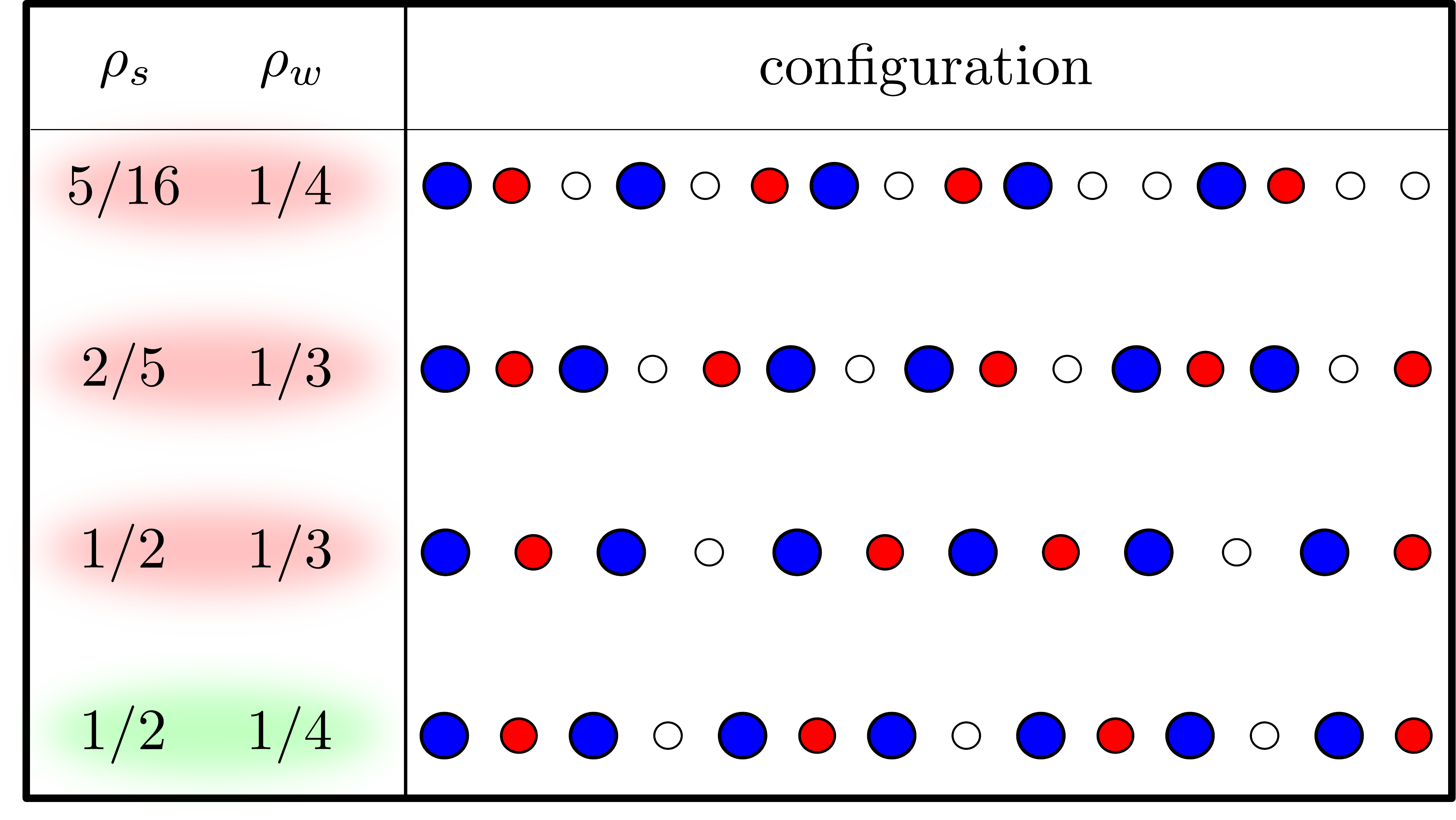}
\end{center}
\caption{(Color online) Microscopic configuration for different choices of the filling fractions $\rho_s$ and $\rho_w$. The excitations of the $s$-species are indicated with solid blue large circles and those of the $w$-species with solid red small circles. Empty sites are represented by empty circles. The case shown at the bottom of the figure is compatible (green shade), such that the arrangement of both the $s$- and $w$-excitations can be achieved by applying the algorithm (\ref{eq:hubbmethod2}) on the two species independently. All other examples are incompatible (red shade).} \label{fig:config}
 \end{figure}
Let us now discuss the actual ground state of the Hamiltonian (\ref{eq:ham2spec}). For compatible filling fractions this is given by the distributions of the two species arranged independently using the algorithm (\ref{eq:hubbmethod2}). In case of the two filling fractions being incompatible the situation is more complicated. In order to make analytical progress we work under the assumption that $V_s/V_w\rightarrow \infty$, suchthat the arrangement of $s$-excitations can be considered as ``frozen''. Note, that such situation can be achieved in the context of a realisation of the system with atomic  Rydberg gases when the two principal quantum numbers corresponding to the $s,w$ states are chosen appropriately, see \cite{Levi_2015} for more details. The problem then reduces to the arrangement of $w$-excitations interacting with a $1/r^\alpha$ potential constrained to sit in the $s$-holes. It is now convenient to define an effective inhomogeneous or distorted lattice with spacings $\tilde{a}\in \left\{\left\lfloor \tilde{\rho}_s^{-1} \right\rfloor ,\left\lceil \tilde{\rho}_s^{-1}\right\rceil \right\}$. Clearly the inhomogeneity alters the minimum energy disposition of the $w$-excitations, such that we cannot consider only nearest neighbor distances $r_1$ and $r_1+1$, but we have in general to consider distances up to $r_1+\beta$, with $\beta>1$. In the following we use the term ``distortion of the minimal set'' to denote that $l$-th nearest neighbours distances can take values in the set $\{r_l, r_l+1,...,r_l+\beta \}$ rather than $\{r_l,r_l+1\}$. The smallest distortion then refers to the case when only one more distance $r_l+2$, i.e. $\beta=2$, is included.

One can now exploit the convexity of the potential to demonstrate that the energy is minimized by the most homogeneous configuration. We start by considering the smallest distortion of the minimal set. Then the following relation holds:
\begin{equation}
 \label{eq:rplus2}
 \sum_{i=1}^{n_w}r_l^{(i)}=(n_w-a-b)r_l+a(r_{l}+1)+b(r_{l}+2)=lN,
\end{equation}
where $a$ and $b$ are the number of occurrences of the distances $r_l+1$ and $r_l+2$ respectively.

We now want to show that the introduction of a longer distance results in an increase of the energy. To this end let us consider the set $\{r_l,r_l+1,r_l+2,r_l+3\}$ where only one additional distance $r_l+3$ is introduced in the ground state configuration, while the total number of excitations $n_w$ is kept fixed.
Then, since the sum (\ref{eq:rplus2}) must be conserved, we have \footnote{Adding the distance $r_l+3$ leads to new number of distances $r_l, r_l+1,r_l+2$ which are constrained to sum up to $Nl$ together with the single distance $r_l+3$. This can be achieved by adding and subtracting $r_l+3$ to (\ref{eq:rplus2}) and absorbing the $-(r_l+3)$ term in the remaining terms. In (\ref{eq:rplus3}) we have just absorbed the $-(r_l+3)$ term in a way that makes the convexity of the potential (\ref{eq:convex}) manifest.}
\begin{equation}
 \label{eq:rplus3}
 \sum_{i=1}^{n_w}r_l^{(i)}=(n_w-a-b-1)r_l+a(r_{l}+1)+(b-1)(r_{l}+2)+(r_l-1)+(r_l+3).
\end{equation}
Calling $E_2$ the energy of the configuration (\ref{eq:rplus2}), and $E_3$ the one of (\ref{eq:rplus3}) the following holds
\begin{equation}
\label{eq:compare2}
E_3-E_2=V(r_l-1)+V(r_l+3)-V(r_l)-V(r_l+2) > 0.
\end{equation}
where the positivity stems from the convexity of the potential (\ref{eq:convex}). Analogously, one can prove by induction that having a larger set of distances is always energetically unfavourable, such that the minimal energy configuration is indeed the one achieved with the minimal set of distances, i.e. with $\beta=2$.

The procedure for finding the ground state configuration in the incompatible case can now be summarized as follows.
\begin{itemize}
 \item Define the inhomogeneous lattice using (\ref{eq:hubbmethod2}) with $\tilde{\rho}_s$.
 \item Distribute the $w$-excitations in the effective lattice given by the disposition of $s$-holes, using algorithm (\ref{eq:hubbmethod2}) with $\rho_w/\tilde{\rho}_s$ as the effective filling fraction.
\end{itemize}
In Fig. \ref{fig:config} we show some examples of the ground state configurations for various filling fractions.

We now want to show that the above-described procedure automatically yields the minimal distortion, i.e. $\beta=2$, of the minimal set of distances of the $w$-excitations. Lets denote the two values assumed by the $s$-hole distances $\tilde{a}$ as $a'$,$a'+1$. It follows from (\ref{eq:hubbmethod2}), that the ground state configuration always contains strings of the form $a'^{k}(a'+1)$ and $a'^{k+1}(a'+1)$, or $(a'+1)^k a'$ and $(a'+1)^{k+1}a'$. The ground state configurations of the $s$-holes can then be divided into four mutually exclusive cases, depending on what kind of string of distances $a'$ they contain. If they contain
\begin{itemize}
 \item $a'^{k+1}(a'+1)a'^{k}(a'+1)a'^{k+1}(a'+1)$,\\
 the maximum distortion occurs for the configurations with $\mathcal{Q}_l=\{k+1,k+2\}$ for $l$th-nearest neighbor
 In this case $s_{q_l}\in\{a'^{k+1},a'^k(a+1)\}$ and $s_{q_l+1}\in\{a'^{k+1}(a'+1),(a'+1)a'^k(a'+1)\}$, such that the maximum difference in length is $\beta=2$.
 \item $a'^{k}(a'+1)a'^{k+1}(a'+1)a'^{k}(a'+1)$,\\
 the maximum distortion occurs for the configurations with $\mathcal{Q}_l=\{k+1,k+2\}$. The maximum difference in length is $\beta=2$.
 \item $(a'+1)^{k}a'(a'+1)^{k}a' (a'+1)^{k}a'$\\
 the maximum distortion occurs for the configurations with $\mathcal{Q}_l=\{k,k+1\}$, giving again $\beta=2$ as the maximum length difference.
  \item $(a'+1)^{k+1}a'(a'+1)^{k}a' (a'+1)^{k+1}a'$,\\
  the maximum distortion occurs for the configurations with $\mathcal{Q}_l=\{k,k+1\}$. Also in this case the maximum difference in length is $\beta=2$.
\end{itemize}
It follows that apart for the compatible cases for which $\beta=1$, the algorithm described above yields automatically the minimal distortion, that is $\beta=2$, and consequently the resulting configuration minimizes the interaction energy.

\section{Stability region of the compatible phase and compatible to incompatible transitions}
\label{sec:stability}
\begin{figure}
 \begin{center}
\includegraphics[width=.5\textwidth]{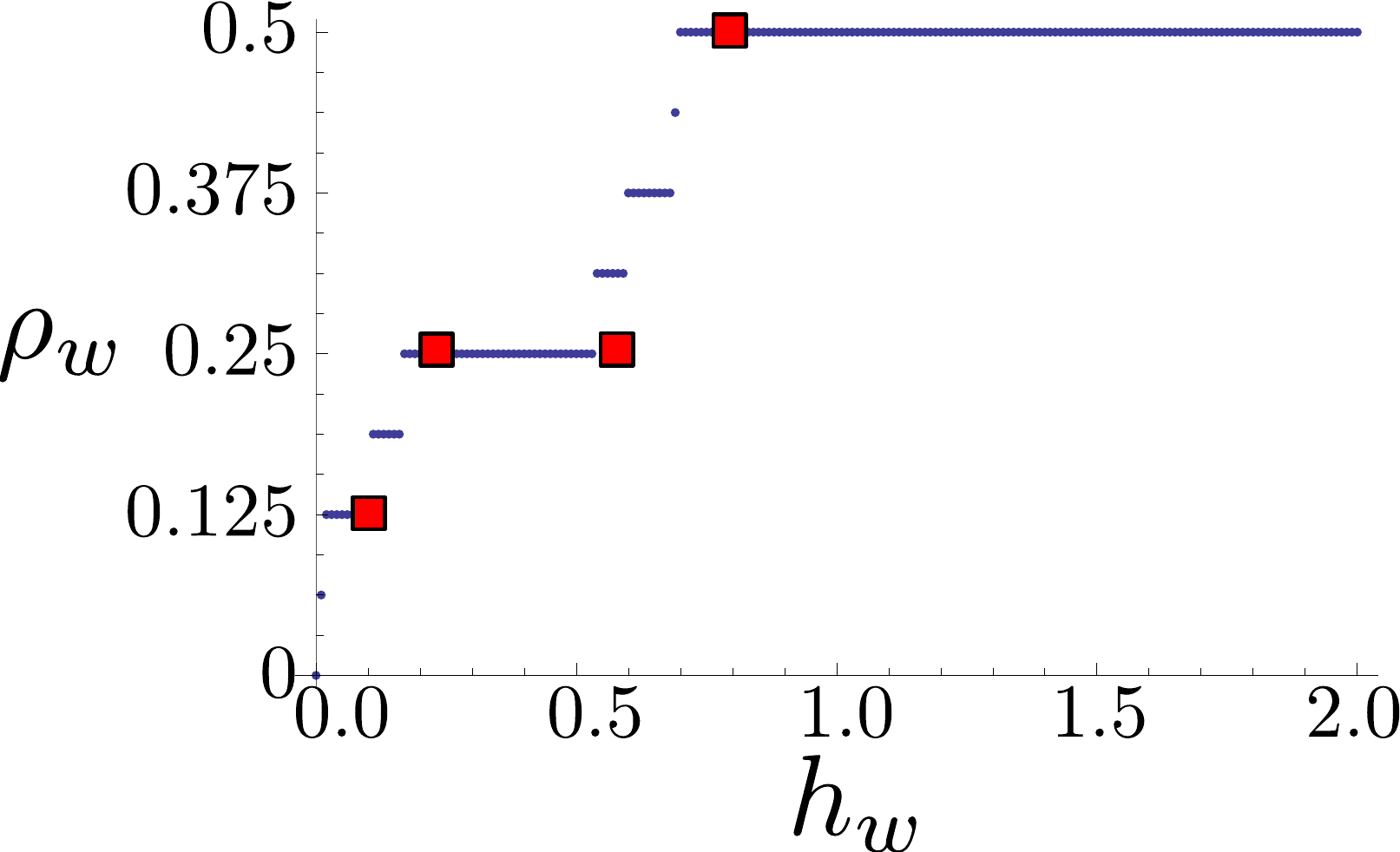}
\end{center}
\caption{(Color online) Stability regions of the $w$-excitations. Numerical results for $\alpha=2$ on a chain of length $N=16$ are shown as blue dots.
The density of $s$-excitations is fixed to $\rho_s=1/2$, and we set the parameters $h_s=V_w=1$ and $V_s=100$. The superimposed red squares are the predictions of Eq. (\ref{eq:stabreg}) for the compatible to incompatible transitions in the thermodynamic limit. } \label{fig:finite}
 \end{figure}
In the case of a single species the convexity of the potential is sufficient to determine the stability of a certain density through Eq. (\ref{eq:system2}). This is generally not true anymore in the two-species case and it is very challenging to determine which configuration or disposition is stable for given values of $h_w$ and $h_s$.

We can make progress by again assuming that $V_s/V_w\rightarrow\infty$. Here the results of the single-species problem apply for the $s$-excitations, while (\ref{eq:rplus2}) holds for the $w$ species, such that e.g. extracting one excitation leads to the following set of equations
 \begin{equation}
  \left\{
   \begin{array}{l}
     (n_w-a-b)r_l+a(r_{l}+1)+b(r_{l}+2)=lN \label{eq:system3_1} \\
     (n_w-a'-b'-1)r_l+a'(r_{l}+1)+b'(r_{l}+2)=lN\\
   \end{array}
   \right. .
 \end{equation}
The aim is to find the variables $a',b'$ in terms of the variables $a,b$, similar to the case of a single species. However, the system of equations (\ref{eq:system3_1}) does not yield a complete solution. Instead it provides the consistency relation
\begin{equation}
  a'-a=r_l-2(b'-b).
  \label{eq:consistency}
\end{equation}

In order to proceed we restrict our analysis to the case $\rho_s=1/q_s$, and $\rho_w=1/q_w$. The independent configurations are clearly homogeneous with periods $q_s$ and $q_w$. The two configurations are incompatible if and only if $q_s$ and $q_w$ have no trivial common divisor. This can be shown as follows. Let us denote the positions of the excitations as $x_s = k q_s$ and $x_w = l q_w + \delta$, with $\delta\in \mathbb{Z}$. This means $\delta$ is a constant shift of the $w$-configuration with respect to the $s$ configuration. If $x_s \neq x_w$ for $\forall l,k \in \mathbb{Z}$, then the two configurations are compatible. We thus have
\begin{equation}
  \delta \neq k q_s - l q_w = p(k q'_s-l q'_w) 
  \label{eq:delta}
\end{equation}
where in the second equality we explicitly collected $p$, the common divisor of $q_\mu$, such that  $q_{\mu} = p q'_{\mu}$, $\mu=s,w$.

The term in brackets at the r.h.s. of (\ref{eq:delta}) is an integer number,
such that for non trivial $p\neq 1$ one can always find a shift $\delta$ (different from multiples of $p$) that yields non-conflicting configurations. On the other hand, if $p=1$, such shift does not exist and the configurations are incompatible.

Particularly interesting is to understand the transition from a compatible disposition (or phase) to an incompatible one. This can be studied as in Eqs. (\ref{eq:sdifferences})--(\ref{eq:region}) by perturbing the minimum energy configuration in the thermodynamic limit by subtracting or introducing one $w$-excitation. If we perturb a compatible configuration (where $q_{s,w}$ have a common divisor $p$) this will with certainty lead to an incompatible disposition, as it will introduce $q_w$ distances $q_w\pm1$:
Every $q_w\pm1$ distance shifts the excitations to the left/right by one site with respect to the ground state configuration (which contains only distances $q_w$ between nearest neighbours). This occurs $q_w$ times and since $q_w>p$, the condition (\ref{eq:delta}) cannot be satisfied.

Let us focus now in detail on the case $\nu=q_w/q_s\in \mathbb{N}^+$ for which the transition points can be found analytically. Using the procedure described in Section \ref{sec:disposition} one finds that the set of $l$-th nearest neighbour distances $\{ lq_w \}$ is modified in the following way. The insertion of one excitation introduces $l\nu$ distances $q_w-2$ and $l(q_w-2\nu)$ distances $q_w-1$. On the other hand subtracting one excitation will introduce $l\nu$ distances $q_w+2$, and $l(q_w-2\nu)$ distances $q_w+1$.
Notice that this is in agreement with the consistency condition dictated by (\ref{eq:consistency}).

Calling $h_w^{(\pm)}$ the values of $h_w$ for which these compatible to incompatible transitions occur, we find
\begin{eqnarray}
\hspace*{-2cm} h_w^{(-)}=\sum_{l=1}^{\infty}\left[\left(1+l(q_w-\nu) \right)V(lq_w)-l(q_w-2\nu)V(lq_w+1)-l\nu V(lq_w+2)\right],\nonumber\\
 \nonumber \\
\hspace*{-2cm} h_w^{(+)}=\sum_{l=1}^{\infty}\left[\left(1-l(q_w-\nu) \right)V(lq_w)+l(q_w-2\nu)V(lq_w-1)+l\nu V(lq_w-2)\right].\nonumber\\
 \label{eq:stabreg}
\end{eqnarray}
Considering a general form of the filling fractions it is very hard to read a repeated pattern of distances, such that we have not been able to determine regions of stability analytically.

In order to verify Eqs. (\ref{eq:stabreg}), we have determined the stability regions numerically for a finite chain of length $N=16$, where we have fixed $\rho_s=1/2$. The results are shown as blue dotted curve in Fig. \ref{fig:finite}. In this particular case the only accessible filling fractions are of the form $m/N$, $m \leq N$. However, since Eqs. (\ref{eq:stabreg}) are only valid for filling fractions of the form $1/q_w$, the only accessible values which can be used for a comparison are $\rho_w=1/8,1/4,1/2$. These regions are delimited by the red squares in Fig. \ref{fig:finite}. We can observe that the predictions in the thermodynamic limit tend to overestimate the values of the external field at which the transition happens compared to the finite chain calculations. This discrepancy is more pronounced for higher density of $w$-excitations. This is understandable as higher filling fractions introduce larger energy corrections in the finite chain.

\section{Conclusions}
We have investigated the statics of a two-component lattice gas with species-dependent $1/r^\alpha$ interactions in a one-dimensional lattice. The motivation behind this study is a link between this system and the current experiments in which cold atoms are excited to multiple high-lying Rydberg states. We found that the ground state arrangement of the species in the lattice falls into one of two possible categories which depend on the filling fractions of the two species. There is a compatible case in which the two species can be considered as independent, and incompatible one in which this is not the case. We have defined the criteria for compatibility of the two species configurations. In the limiting case in which we consider one of the two species as frozen, we showed how to determine the ground state configuration in the incompatible phase. Finally, for filling fractions of the form $1/q$, we determined the stability regions of the compatible phases in the thermodynamic limit and compared the analytical result with finite size numerical simulation.

\section{Acknowledgements}
E.L. would like to thank M. Marcuzzi for insightful discussions. The research leading to these results has received funding from the European Research Council under the European Union's Seventh Framework Programme (FP/2007-2013) / ERC Grant Agreement No. 335266 (ESCQUMA), the EU-FET grants HAIRS 612862 and from the University of Nottingham.  Further
funding was received through the H2020-FETPROACT-2014 Grant No.  640378 (RYSQ) and the EPSRC Grant no.\ EP/M014266/1.
\\
\\
\bibliographystyle{iopart-num}

\providecommand{\newblock}{}

\end{document}